# Topology Classification from Chiral Symmetry: Chiral Phase Index and Spin Correlations in Graphene Nanoribbons

Jingwei Jiang[1,2] and Steven G. Louie[1,2,*]

[1]Department of Physics, University of California, Berkeley, CA 94720, USA.

[2]Materials Sciences Division, Lawrence Berkeley National Laboratory, Berkeley, CA 94720, USA.

*correspond to: sglouie@berkeley.edu

**Abstract**

Topology concepts have significantly deepened of our understanding in recent years of the electronic properties of one-dimensional (1D) nano structures such as the graphene nanoribbons[1]. Controlling topological electronic properties of GNRs has been demonstrated in both theoretical studies[1,2,3] and experimental realization[4,5]. Most previous works rely on classification theory requiring both time reversal and spatial symmetry of a unit cell in the 1D bulk material that is commensurate to its boundary. To access boundary structures that lead to unit cell with no spatial symmetry and to generalize the theory, we propose here another classification scheme, using chiral symmetry, to arrive at a Z classification that is not only applicable to GNRs with arbitrary terminations, but also to any general 1D chiral structures. This theory, combining with Lieb's theorem[6], moreover enables access to the electron's spin degree of freedom, allowing for investigation of spin physics.

**Main**

Topology classification theory has broadly been applied to explain many physical phenomena such as quantum Hall insulators[7,8,9,10,11], quantum spin Hall insulators[12,13], topological insulators and superconductors[14,15,16,17,18]. The power of topology theory has not been as widely used in one-dimensional systems. The recently developed bottom-up molecular precursors technique enables the synthesis of atomically-precise graphene nanoribbons (GNRs)[19,20,21,22]. These structurally precise 1D materials have been predicted to possess band gap due to quantum confinement[23,24] while graphene is a well-known semi-metal. Since the discovery of distinct topological phases in GNRs[1], topology classification in GNRs has proven to be highly successful in predicting the emergence of topological in-gap states localized at the boundary of such GNRs. However, the origin of the observed robust junction states between bearded termination of GNRs[5] and metallic armchair graphene nanoribbons (AGNRs) formed by superlattice of topological in-gap states[25] is still unclear. In the former kind of junction (Fig.1 c)), as spatial symmetry in the commensurate unit cell on both sides is no longer preserved, the previous $Z_2$ topology theory based on spatial symmetry loses its prediction power. To overcome this conceptual issue, we develop in this paper an approach using chiral symmetry instead to classify 1D structure. Crystals in the honeycomb structure have chiral symmetry if second-nearest-neighbor interaction can be neglected.

Mathematically, chiral symmetry (sublattice symmetry) is a parity (P-type) symmetry acting on the sublattice index of the Hamiltonian $H$, satisfying[17]:

$$H = -PHP^{-1},$$

$$PP^\dagger = 1,$$

$$P^2 = 1 \qquad (1)$$

The operation does not depend on any spatial coordinates and thus can easily be preserved when a bulk system is terminated at a boundary. Chiral symmetry is exact in bipartite lattices, in which the system can be divided into two sets of sublattice A and B, such that interactions only exist between atoms on different sublattices. For such a system, Eq. (1) is satisfied by setting $P$ to a diagonal matrix with 1 for the A sublattice part and -1 for the B sublattice part. Graphene is a bipartite lattice system within a tight-binding formalism, with only first-nearest-neighbor hopping included. Within this spirit, we may analyze the electron topological properties of GNRs, and any other 1D structures with nearly bipartite symmetry, using chiral symmetry and treat subsequently the small second-nearest-neighbor effects on the results (such as the in-gap boundary states) perturbatively.

We now apply this idea to the AGNRs by setting proper boundary conditions in the transverse direction of AGNRs. To derive the bulk-index for the AGNRs, we use the first-nearest-neighbor tight-binding model and follow the standard Fermion-projector methods[18]. The Fermion-projector, written in reciprocal space, is a Hermitian operator defined as:

$$Q_k = \sum_{n}^{N_{unocc}} |\psi_{nk}\rangle\langle\psi_{nk}| - \sum_{m}^{N_{occ}} |\psi_{mk}\rangle\langle\psi_{mk}| \quad (2)$$

Where $|\psi_{nk}\rangle$ stands for Bloch states of band n and momentum k. $N_{unocc}$ represents the number of unoccupied bands and $N_{occ}$ is the number of occupied bands. $Q_k$ can be understood as a continuous deformation of the original Hamiltonian $H$ with a gapped spectrum by moving every eigen energies of the occupied bands to be -1 and unoccupied ones to be +1, while keeping the eigenvectors unchanged. Using the normalization, orthogonality, and completeness of the eigenvectors of H, one can show that $Q_k^2 = 1$. Under chiral symmetry, $Q_k$ could be brought into off-diagonal form using localized site basis:

$$Q_k = \begin{bmatrix} 0 & U_k \\ U_k^\dagger & 0 \end{bmatrix} \quad (3)$$

Combined with above properties, it is proved that $U_k$ belongs to unitary group $U(N_{occ})$, the classification is given by homotopy group $\pi_1(U(N_{occ})) = Z$ and the bulk-index is written as[26]:

$$Ch_1(U) = \frac{-i}{2\pi} \int_{1D\ BZ} Tr(U_k^\dagger \partial_k U_k)\, dk \quad (4)$$

This is equivalent to the winding of the phase of the determinant of $U_k$ over the closed path of 1D Brillouin zone (see supplementary for details). Although this formula has been broadly used in mathematics, it is inconvenient to use the matrix form to evaluate physical quantities. Here, we use the properties of wavefunctions under chiral symmetry to further simplify the bulk-index into vector form, which we shall name as the chiral phase index (CPI).

From Eq. (1), the Hamiltonian with chiral symmetry anticommute with the chiral operator $P$ in a bipartite lattice. This brings the following properties for the wavefunctions:

$$P \begin{pmatrix} \alpha_{nk} \\ \beta_{nk} \end{pmatrix}_E = \begin{pmatrix} \alpha_{mk} \\ \beta_{mk} \end{pmatrix}_{-E} = \begin{pmatrix} \alpha_{nk} \\ -\beta_{nk} \end{pmatrix}_E. \quad (5)$$

The subscript $E/-E$ represents the eigenvalue of Hamiltonian. $\alpha_{nk}, \beta_{nk}$ are vectors representing A sublattice components and B sublattice components respectively. Combining Eq. (2-5), and using the orthogonalization properties of wavefunctions (see supplementary for details), we derive the chiral phase index (CPI):

$$Z = \frac{-i}{\pi} \sum_{n \in occ} inter \left( \int_{1D\ BZ} \langle u_{nk}|P\partial_k|u_{nk}\rangle dk \right) \quad (6)$$

*inter* means taking only the intercell part[27]. $u_{nk}$ is the periodic part of the Bloch states. This new formula of bulk-index, expressing as a function of eigenvectors, requires only the knowledge of wavefunctions from a tight-binding framework, rather than full information of the Hamiltonian matrix and is very convenient in calculate bulk-index analytically after the eigenvectors are obtained. Before evaluating the bulk-index of AGNRs, we want to point out that the CPI in our work is different with the intercell part of Zak phase used in previous work[1,28,29], namely $\sum_{n \in occ} inter \left( \int_{1D\ BZ} \langle u_{nk}|\partial_k|u_{nk}\rangle dk \right)$. CPI evaluates the *difference* between intercell part of Zak phase contributed by A sublattice and that contributed by B sublattice rather than the sum of these two parts, and yields Z classification rather than $Z_2$. An important consequence is that CPI is fully gauge-invariant while intercell Zak phase is only invariant mod $2\pi$. This can be shown by expanding CPI into A,B components explicitly, using expression in Eq. (5):

$$Z = \frac{-i}{\pi} \int_{1D\ BZ} dk \sum_{n \in occ} (\alpha_{nk}^\dagger \partial_k \alpha_{nk} - \beta_{nk}^\dagger \partial_k \beta_{nk}) \quad (7)$$

Apply a gauge transform: $\alpha_{nk} \to \alpha_{nk} e^{if_n(k)}$ and $\beta_{nk} \to \beta_{nk} e^{if_n(k)}$, the CPI changes to:

$$Z = \frac{-i}{\pi} \int_{1D\ BZ} dk \sum_{n \in occ} (\alpha_{nk}^\dagger \partial_k \alpha_{nk} - \beta_{nk}^\dagger \partial_k \beta_{nk})$$

$$+ \frac{1}{\pi} \int_{1D\ BZ} dk \sum_{n \in occ} \left( \alpha_{nk}^\dagger \alpha_{nk} \partial_k f_n(k) - \beta_{nk}^\dagger \beta_{nk} \partial_k f_n(k) \right) \quad (8)$$

$$= \frac{-i}{\pi} \int_{1D\ BZ} dk \sum_{n \in occ} (\alpha_{nk}^\dagger \partial_k \alpha_{nk} - \beta_{nk}^\dagger \partial_k \beta_{nk}) + 0 \quad (9)$$

From Eq. (8) to Eq. (9), we have used to relation in supplementary Eq. (6). This gauge invariant property is an essential character of the CPI, leading to its value being integer numbers, in contrast with the Zak phase which could only be 0 and $\pi$ ($Z_2 = 0, or\ 1$) under time reversal and spatial symmetry.

We note here another important distinct between the new CPI and the Zak phase. As formulated, the CPI is only defined for a system with Fermi level in the charge neutrality gap since chiral symmetry in the form of Eq. (3) has been use.

As a gauge invariant quantity, the CPI is expected to be a measurable quantity. While Zak phase in one dimension connects with the modern theory of polarization[30], the CPI is related to the difference between the electric dipole moments per unit cell of the two sub-lattices. However, the most straightforward way of measuring CPI would be counting the topological edge states at the edge of the system with vacuum. This quantity is connected with the bulk-index by using the bulk-edge correspondence given as[26]: $Z_{bulk} = N_+ - N_-$ when the system is terminated to the right. $N_{+(-)}$ is the number of zero-mode in the charge neutrality gap with positive (negative) chirality. States with positive chirality will localize only at the A sublattice while states with negative chirality localizes only at the B sublattice. In summary, through bulk-edge correspondence, the CPI contains two pieces of important information for the terminated system. Firstly, $|Z|$ gives the number of edge states that is protected by symmetry; and secondly, $sgn(Z)$ gives the chirality of the edge states. When the system is terminated to the left, we could obtain the bulk-edge correspondence by rotating the bulk material by 180°, this takes $u_{nk} \rightarrow u_{n,-k}$. Substituting into Eq. (6), we get: $Z_{bulk} = N_- - N_+$.

Moreover, the bulk-edge correspondence applies when two such bulk materials with distinct CPI are joint and symmetry-protected junction states are formed. One can show that the number of symmetry-protected topological junction states, as two bulk structures with bulk index $Z_{bulk}^{left}$ and $Z_{bulk}^{right}$ are joined, is $N_+^{junc} - N_-^{junc} = Z_{bulk}^{left} - Z_{bulk}^{right}$.

We now obtain an explicit expression for the CPI of AGNRs with different widths and terminations using Eq. (7). The tight-binding wavefunctions of the GNRs may be analytically calculated from a linear combination of graphene's wavefunctions with proper boundary conditions[1] (see supplementary). The orthogonalization and completeness of wavefunctions are proved using the orthogonalization of graphene's Bloch wavefunctions and the dimension of the Hilbert space. The gauge for the Bloch wavefunctions is chosen such that it is continuous along $k$ and satisfies periodic boundary condition across the 1st Brillouin zone. After inserting the resulting wavefunctions into Eq. (7), a general form of Z index is derived as

$$Z = N_{notco} - \left\lfloor \frac{N}{3} \right\rfloor \qquad (10)$$

$N$ is the total number of rows of carbon atoms forming the width of the AGNR[31] and $N_{notco}$ is the number of rows of atoms with carbon pairs not connected by $\sigma$ bonds within the unit cell. The topless brackets denote the floor function which takes the largest integer less than or equal to the value in the brackets. The definition of connected carbon pairs (with distance close to $a/\sqrt{3}$) and unconnected pairs (with distance close to $\sqrt{3}a/2$ ) are shown in Fig. 1**b**). Fig. 1**b**) also illustrates the convention of defining A and B sublattice in AGNR. This convention will be followed throughout the paper.

With the Z classification using chiral symmetry, the junction states localized at the asymmetric junction in Fig. 1 formed by bearded termination of 7-AGNR and 9-AGNR observed in experiment[5,25], could now be well explained. The bearded 7-AGNR unit cell has $Z = -2$, while 9-AGNR unit cell has $Z = -3$, causing one in-gap junction state to arise at the junction, as confirmed by density functional theory (DFT) calculations within the local-density approximation (LDA). Since $Z_{bulk}^{left} - Z_{bulk}^{right} = +1$ , the junction state has amplitudes only on the A sublattice (Fig. 1**c**)).

The theory can also be applied to more general termination types of AGNR, not limited to zigzag, zigzag', or bearded types[1]. As shown in Fig. 2**a**), the zigzag termination of 7-AGNR has $Z = -1$ and "bullet" termination of 9-AGNR has $Z = 1$. When terminated to the vacuum, one edge state shows up at the termination of each structure. Nevertheless, the two corresponding bulk structures still belong to difference classes because of the opposite sign of the CPI. Two topologically protected states should occur when the two structures are joint, as confirmed by DFT-LDA calculation (Fig. 2**b**)). A simple understanding why these two in-gap localized states do not hybridize significantly with each other and move out of the gap is that they are localized on the same sublattice. Since by construction the whole system preserves chiral symmetry, there is no interaction which would hybridize them -- due to the fact that any interaction at the junction allowing hopping between same sublattice sites would require broken chiral symmetry. In principles, for the real AGNR junction, there may be a small energy splitting between the two junction states due to second-nearest-neighbor hopping. Our DFT-LDA results show that such splitting due to breaking of chiral symmetry in this case is minimal.

Furthermore, if the spin degree of freedom is considered, the two junction states depicted in Fig. 2b) would couple to each other ferromagnetically if one can arrange for the atomic structure of such a junction to have locally a sublattice imbalance of two carbon atoms, according to Lieb's theorem[6]. And if such a junction is repeated into a 1D superlattice, an 1D ferromagnetic spin chain would form. The insert in Fig. 3a) illustrates the unit cell of such a superlattice. Since each superlattice unit cell has a sublattice imbalance of two, we have a net magnetization of two Bohr magnetons per unit cell. We have confirmed this by performing a DFT-LSDA calculation, and the magnetization is mainly contributed by the two occupied symmetry-protected junction states (Fig .3), in agreement with the conclusion predicted by our topology theory and Lieb's theorem. Since the exchange coupling $J$ between spins is proportional to their wavefunction density overlap, having the two states at mainly at either side of the same junction is expected to have a strong local coupling. To analyze the magnetic properties of such a chain, we map the problem to a 1D Heisenberg model Hamiltonian with two spins per unit cell:

$$H = \sum_i J_1 \hat{s}_{i1} \cdot \hat{s}_{i2} + J_2 \hat{s}_{i2} \cdot \hat{s}_{i+1,1} \qquad (11)$$

where $i$ denotes the unit cell index. To extract the coupling strengths, $J_1$ and $J_2$, from first-principles calculations, we consider three different spin configurations shown in Fig.4 **b)** and performed DFT-LSDA studies as a meanfield level approximation to the Heisenberg model. The first configuration corresponds to a state with total energy $\frac{1}{4}J_1 + \frac{1}{4}J_2$ per unit cell. The second configuration has energy of $-\frac{1}{4}J_1 - \frac{1}{4}J_2$, and the last one has energy of $\frac{1}{4}J_1 - \frac{1}{4}J_2$. The total energy differences from our first-principles calculations under LSDA may be used to extract the exchange coupling strength parameters, which yield $J_1 = -87\ meV/\hbar^2$ and $J_2 = -30\ meV/\hbar^2$, making them parameters for a stronger ferromagnetic (FM) 1D systems compared to what has been achieved before[32] . Since exchange coupling decays exponentially along distance[1], we estimate the second nearest neighbor exchange to be around $-6\ meV/\hbar^2$ and can be ignored in our Heisenberg model.

Next, we would like to investigate the robustness of the magnetic order. In practice, most GNRs are synthesized on gold substrate, doping and hybridization/screening effects of gold tend to reduce the magnetic order[25].  We found that, within LSDA at T=0, the FM order remains stable up to a transfer of 1.5 electrons per unit cell into the system (see supplementary). Another effect that would reduce FM order is thermodynamic fluctuations. It is known that there is no long-range magnetic order at finite temperature in 1D structures with isotropic spin interactions, according to Mermin-Wagner theorem[33]. Thus, the meaningful quantity should consider is the spin-spin correlation length. As a rough estimate of this quantity, we may use a canonical ensemble of 1D Ising model:

$$H = \sum_i J_1 s_{1i} s_{2i} + J_2 s_{2i} s_{1,i+1} \text{ with } s_{\alpha i} = \pm \frac{1}{2}\hbar \quad (12a)$$

$$Z = \sum_{\{s_{\alpha i}\}} e^{-\beta H(\{s_{\alpha i}\})} \quad (12b)$$

The spin-spin correlation length, defined as $a = -R/\ln\langle s_{\alpha i} s_{\alpha,i+R}\rangle$, where $R$ is the distance between two spins, can be calculated analytically treating the spins classically. Evaluate the expectation value and express it as a function of coupling strength and temperature, we have[34]:

$$a = -\frac{2}{\ln(|\tanh \beta J_1|) + \ln(|\tanh \beta J_2|)}, \quad (13)$$

where $\beta = 1/k_B T$. The temperature dependence of correlation length is plotted in Fig. 4**c**). At 3K (at which low temperature STM measurements typically are done), the spin correlation is expected to be at tens-of-nanometers scale. The strong coupling strength and long correlation length of such designed GNR should open up applications to spin qubits[35] and spin-dependent transport[36] through nanostructures.

As a final remark, we would like to point out that the present classification theory could be applied generally to generate a variety of spin configurations. One could design junctions with arbitrary numbers of coupled localized spin states, and by controlling how the junctions are connected, either FM or AFM coupling between each junction could be realized. The theory may also be applied to other 1D chiral structures, such as the 1D chiral GNRs[37].

## Method

First-principles DFT calculations in the local-density approximation and local spin-density approximation are done using Quantum Espresso packages[38]. A 15 Å vacuum level is applied to nonperiodic (normal to the carbon plane) direction of each material. The atomic geometry of the junction and spin chain structure is fully relaxed until every components of the forces on each atom are smaller than 0.01 eV/Å. Scalar relativistic and norm-conserving pseudo potentials of C and H are used[1,38].

## Acknowledgement

We acknowledge helpful discussions with F. Zhao, Z.L. Li., M. Wu. and T. Cao. This work is supported by Office of Naval Research MURI under award No. N00014-16-1-2921(topology theory), and by the National Science Foundation DMR-1926004(LSDA simulations and spin physics analysis). Computational resources were provided by the DOE at Lawrence Berkeley National Laboratory's NERSC facility and the NSF through XSEDE resources at NICS.

## Author contributions

S.G.L. conceived and directed the research, and J.J. developed the chiral topological formalism and carried out the calculations.

## Corresponding authors

Correspondence to Steven G. Louie at sglouie@berkeley.edu

## Ethics declarations

The authors declare no competing interests.

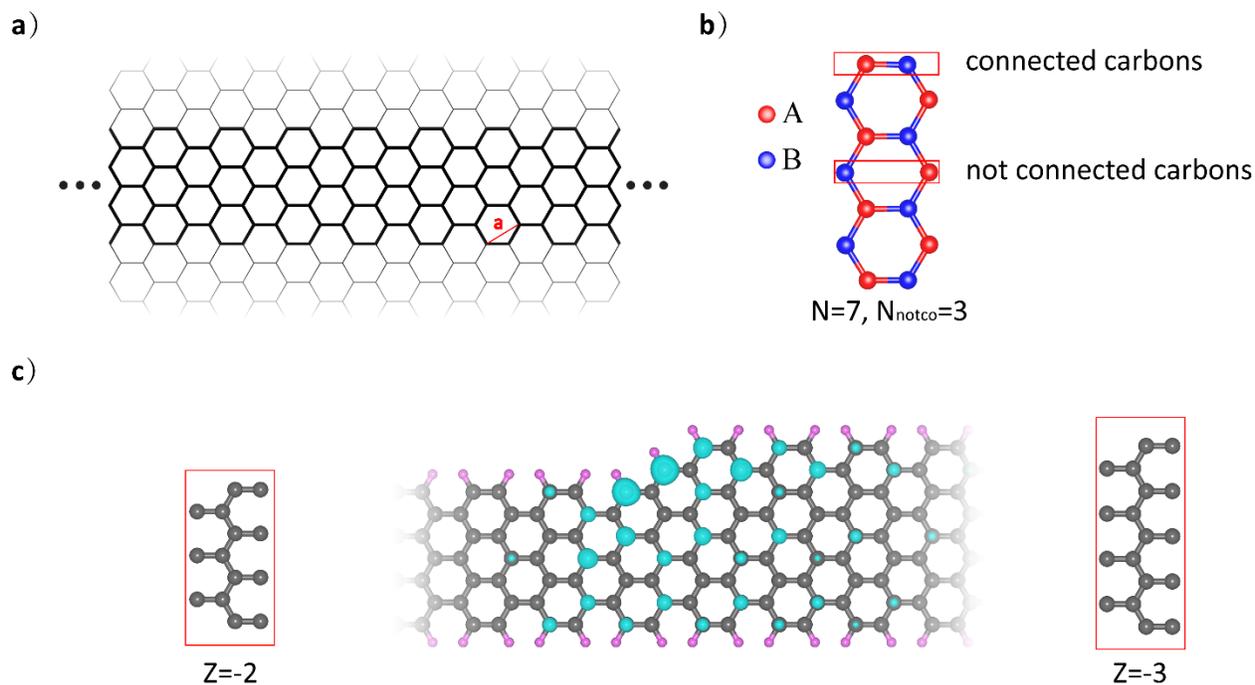

*Figure 1. **a**) the structure of 1D AGNR (bold region) from the graphene backbone background. a is the length of lattice vector of graphene. **b**) unit cell of 7-AGNR with zigzag termination, the rows with two carbon connected by σ bonds within the unit cell (connected carbons) and the rows with two carbon not connected by σ bonds within the unit cell (not connected carbons) are indicated. This case corresponds to having 3 rows of not connecting pairs, $N_{notco} = 3$. **c**) an asymmetric junction of 7-AGNR and 9-AGNR with bearded termination. The corresponding commensurate unit cells of the bulk AGNR for the two segments are shown on the sides, and the 5% isosurface of the junction state from the DFT-LDA calculation is shown in the middle.*

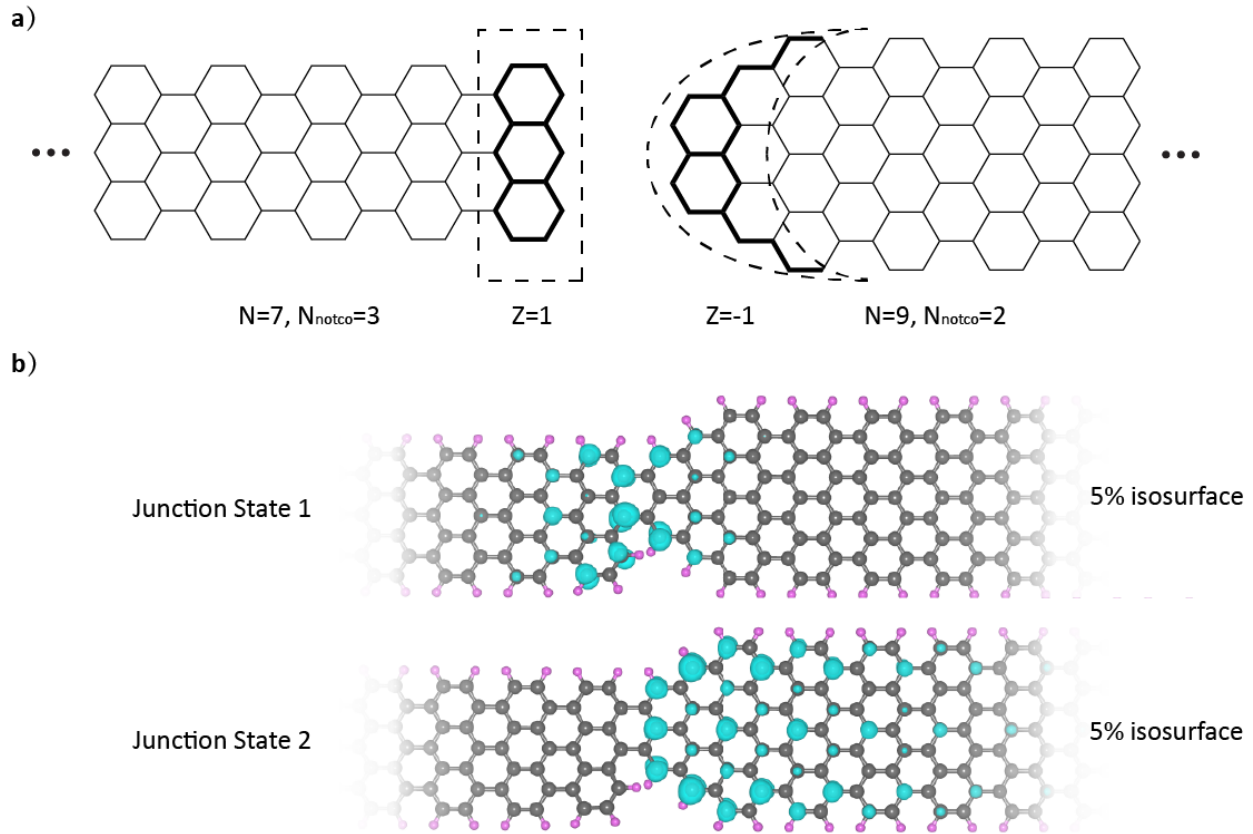

Figure 2. a) 7AGNR with zigzag termination is shown on the left; the unit cell commensurate with the termination has 3 rows of unconnected carbons pairs and Z=1. On the right, a "bullet" termination of 9AGNR is shown. Its commensurate unit cell has 2 rows of unconnected carbons pairs and Z=-1. b) Joining the two structures in a) results in $\Delta Z = 2$, giving rise to two in-gap junction states. The 5% isosurface of the wavefunction square of the two junction states from a DFT-LSDA calculation are shown in blue. Here only the occupied spin-up states are shown. Spin-down states are identical (see supplementary material). One of them localizes in 7AGNR region and the other localizes in 9AGNR region.

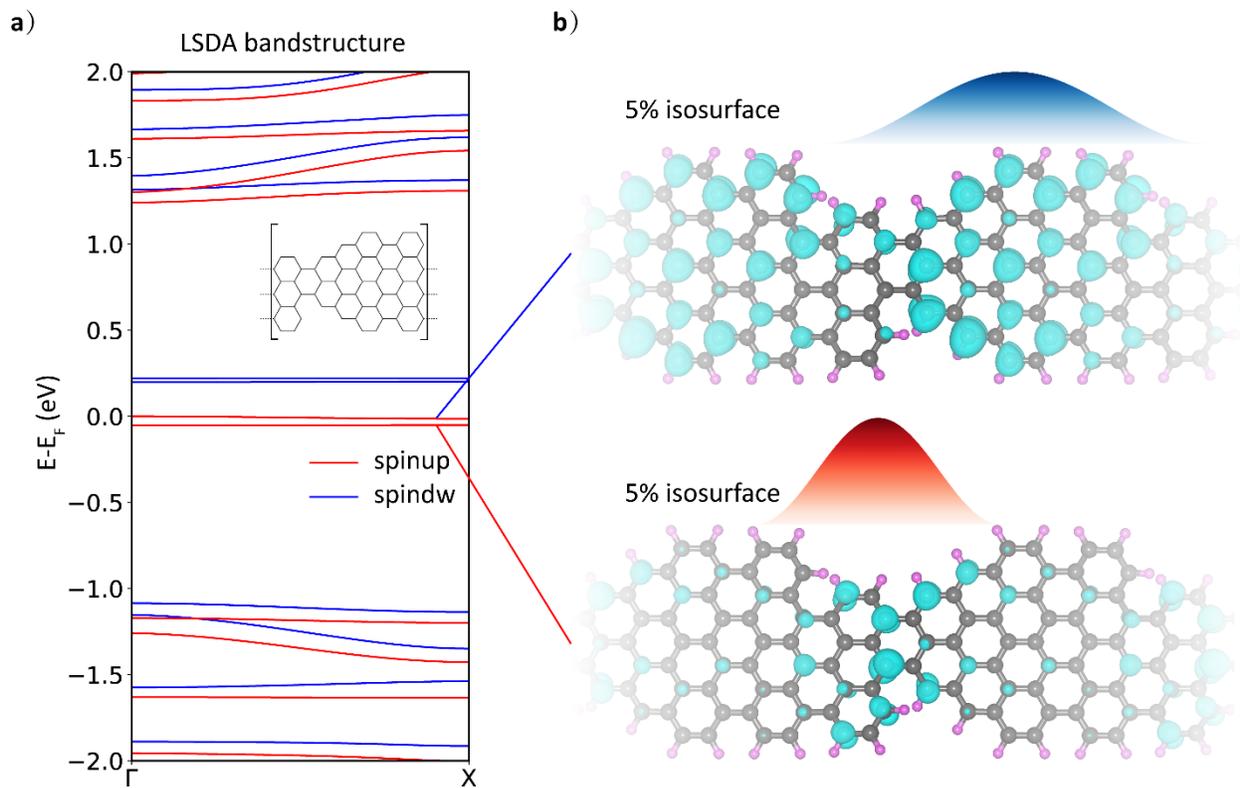

Figure 3 **a)** Computed DFT bandstructure of a periodic GNR spin chain structure (unit cell shown by insert) in the LSDA approximation. The in-gap bands are flat, indicating negligible hopping between junction states of neighboring unit cells, a spin-splitting of 0.2 eV occurs between the spin-up and spin-down bands, and the two spin-up bands are occupied, leaving their spin-down counterparts empty. In agreement with Lieb's theorem. Each unit cell has two Bohr magnetons of magnetization. **b)** The isosurface at 5% of the wavefunction square of the two occupied junction states at k=Γ is shown. One of them is localized in the 9AGNR region while the other is localized in the 7AGNR region.

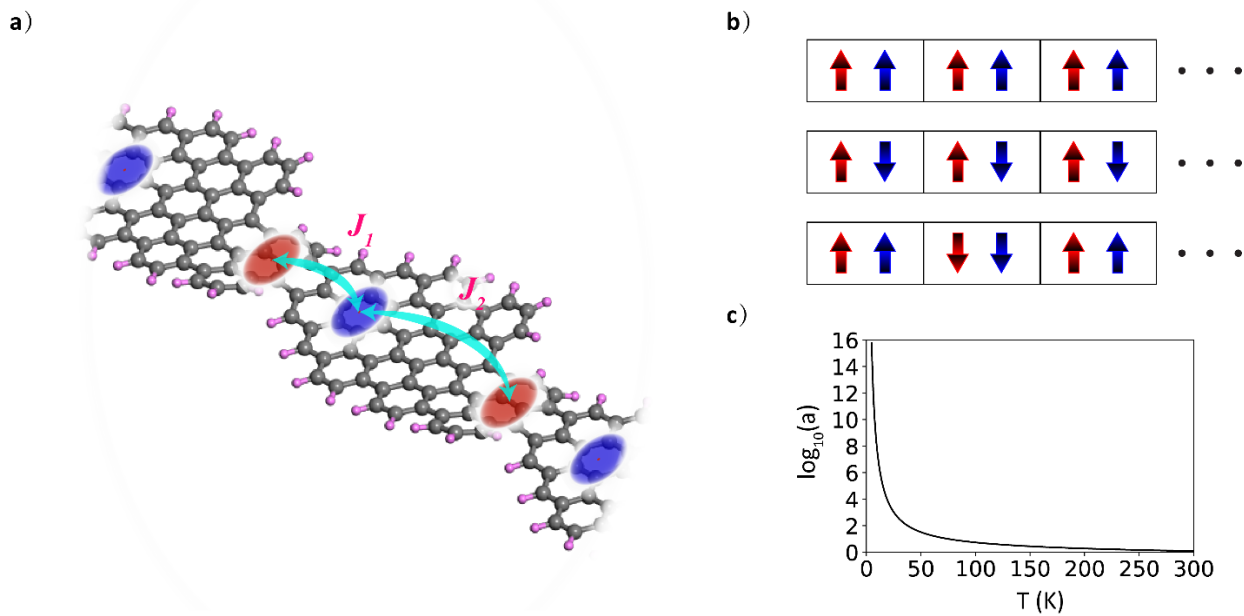

Figure 4 *a*) Schematic of an 1D GNR spin chain and exchange interactions ($J_1$ and $J_2$). *b*) three different spin configurations are considered in the first-principles DFT-LSDA calculations used to extract the exchange coupling strength parameters. *c*) logscale spin-spin correlations in unit of lattice vector as a function of temperature from canonical ensemble of 1D Ising model.